\documentclass[runningheads]{llncs}

\usepackage{graphicx}
\usepackage{caption}
\usepackage{subcaption}
\usepackage[utf8]{inputenc}
\begin{document}
\title{GeneNet VR: Interactive visualization of large-scale biological networks using a standalone headset}
\titlerunning{GeneNet VR}

\author{Álvaro Martínez Fernández\inst{1} \and
Lars Ailo Bongo\inst{2} \and
Edvard Pedersen\inst{2}}

\authorrunning{A. M. Fernández et al.}

\institute{Ramsalt Lab \footnote{Work performed while at UiT - The Arctic University of Norway} \\
\email{alvaro@ramsalt.com} \and 
Department of Computer Science, UiT - The Arctic University of Norway 
\email{lars.ailo.bongo@uit.no, edvard.pedersen@uit.no}}

\maketitle

\begin{abstract}

Visualizations are an essential part of biomedical analysis result interpretation. Often, interactive networks are used to visualize the data. However, the high interconnectivity, and high dimensionality of the data often results in information overload, making it hard to interpret the results. To address the information overload problem, existing solutions typically either use data reduction, reduced interactivity, or expensive hardware. We propose using the affordable Oculus Quest Virtual Reality (VR) headset for interactive visualization of large-scale biological networks.

We present the design and implementation of our solution, GeneNet VR, and we evaluate its scalability and usability using large gene-to-gene interaction networks. We achieve the 72 FPS required by the Oculus’ performance guidelines for the largest of our networks (2693 genes) using both a GPU and the Oculus Quest standalone. We found from our interviews with biomedical researchers that GeneNet VR is innovative, interesting, and easy to use for novice VR users.

We believe affordable hardware like the Oculus Quest has a big potential for biological data analysis. However, additional work is required to evaluate its benefits to improve knowledge discovery for real data analysis use cases.

GeneNet VR is open-sourced: \url{https://github.com/kolibrid/GeneNet-VR}.
A video demonstrating GeneNet VR used to explore large biological networks: \url{https://youtu.be/N4QDZiZqVNY}.
\keywords{VR  \and Visualization \and Graph}
\end{abstract}

\section{Introduction}

The cost of producing data has plummeted in recent years including in  genomics where hundreds of humans can be sequenced in a few days with a cost of around \$1,000 per genome\cite{big_biological_impacts_bd}. However, the interpretation of this vast amount of data requires novel visualization tools to find and understand novel patterns in the data that may lead to new scientific discoveries. In bioinformatics, an often used visualization is a network where the nodes or bioentities are connected by different associations such as gene-to-gene interaction. The networks can be large with 100 thousands of nodes and millions of edges. For these, many of the existing visualization systems for biological networks do not scale or the user interactions are cumbersome. Virtual Reality (VR) enables new ways to implement tools for interactive data visualization \cite{biovr}. However, to use VR to reduce the information overload problem, requires understanding and overcoming the scalability limitations for such biological networks to achieve the required 72 Frames Per Second (FPS) performance for smooth interaction.

\begin{figure}
\includegraphics[width=\textwidth]{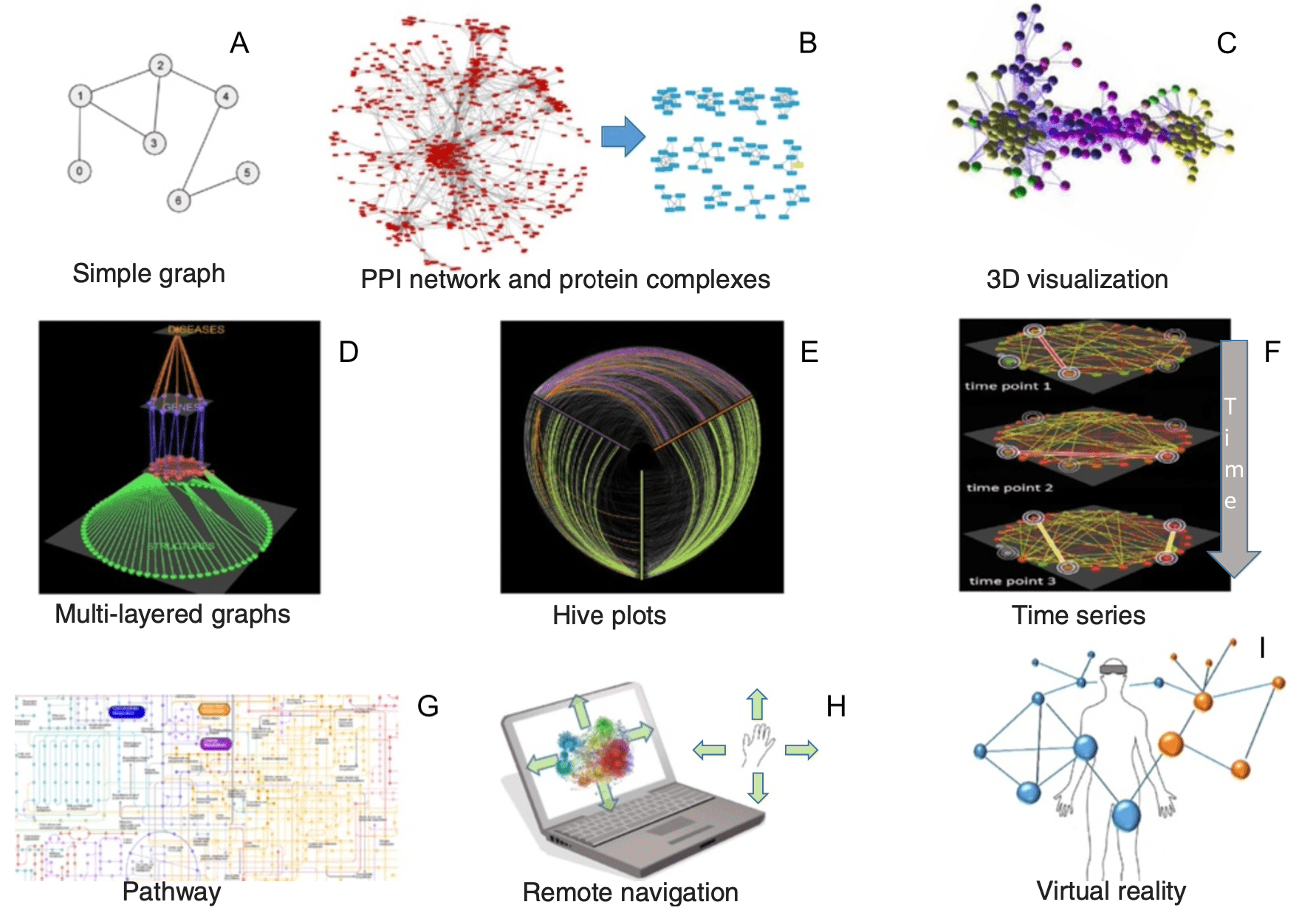}
\caption{Visualization for network biology. (a) Undirected unweighted graph showing co-expression relationship between genes. (b) A 2D representation of a yeast protein-protein interaction network visualized in Cytoscape (left) and potential protein complexes 3D identified by the MCL algorithm from that network (right). (c) A 3D network of genes showing co-expression relationships. (d) A multilayered network integrating different types of data visualized by Arena3D. (e) A hive plot view of a network where nodes are mapped to and positioned on radially distributed linear axes. (f) Visualization of network changes over time. (g) Static picture showing part of lung cancer pathway. (h) Navigation of networks using hand gestures. (i) Integration and control of 3D networks using VR devices. Figure adapted from \cite{pavlopoulos_malliarakis_papanikolaou_theodosiou_enright_iliopoulos_2015}} \label{fig:evolution}
\end{figure}

Existing visualization approaches for large biological networks have three main limitations (Figure \ref{fig:evolution}). First, many are 2-dimensional. Although some provide a 3-dimensional view, they are designed for use on 2-dimensional screens and therefore do not utilize the 3-dimensional visualization and interaction possibilities of  VR. Therefore they often get the "hairball" effect when the network becomes larger. Second, tools limit interactivity to show more data \cite{agapito_guzzi_cannataro_2013}. Third, they need to use specific hardware (BioLayoutExpress3D)\cite{biolayout3d} in order to improve the response time when the number of nodes increases.

We have implemented GeneNet VR, a virtual reality application for the visualization of large biological networks. We used two datasets from the MIxT bioinformatics application \cite{dumeaux_fjukstad_interactions_tumor_blood} containing gene expressions from the NOWAC breast cancer study \cite{nowac}. MIxT provides a 2-dimensional visualization tool to explore gene-to-gene interaction in these datasets. We improve the MIxT visualizations by using a three-dimensional immersive space to implement interactive visualizations. We provided design and implementation guidelines to achieve the necessary scalability and performance.  We implemented and evaluated these on Oculus Quest, a popular and cheap Virtual Reality headset.

We evaluated the performance of GeneNet VR and the interactions mostly used in network exploration. We compare the performance and scalability of the Oculus Quest with and without an external GPU. We found  that GeneNet VR performs well for the network sizes used in MIxT and that the interactions with the network achieved the recommended 72 FPS for VR applications. We concluded that we can use inexpensive VR hardware to explore large biological networks.

We conducted  semi-structured interviews with six biomedical scientific researchers. The feedback that we obtained was positive, highlighting that the application is helpful for the visualization of biological networks and easy to learn even for novice VR users.

\subsection{Use case: Matched Interaction Across Tissues (MIxT)}
\begin{figure}
\includegraphics[width=\textwidth]{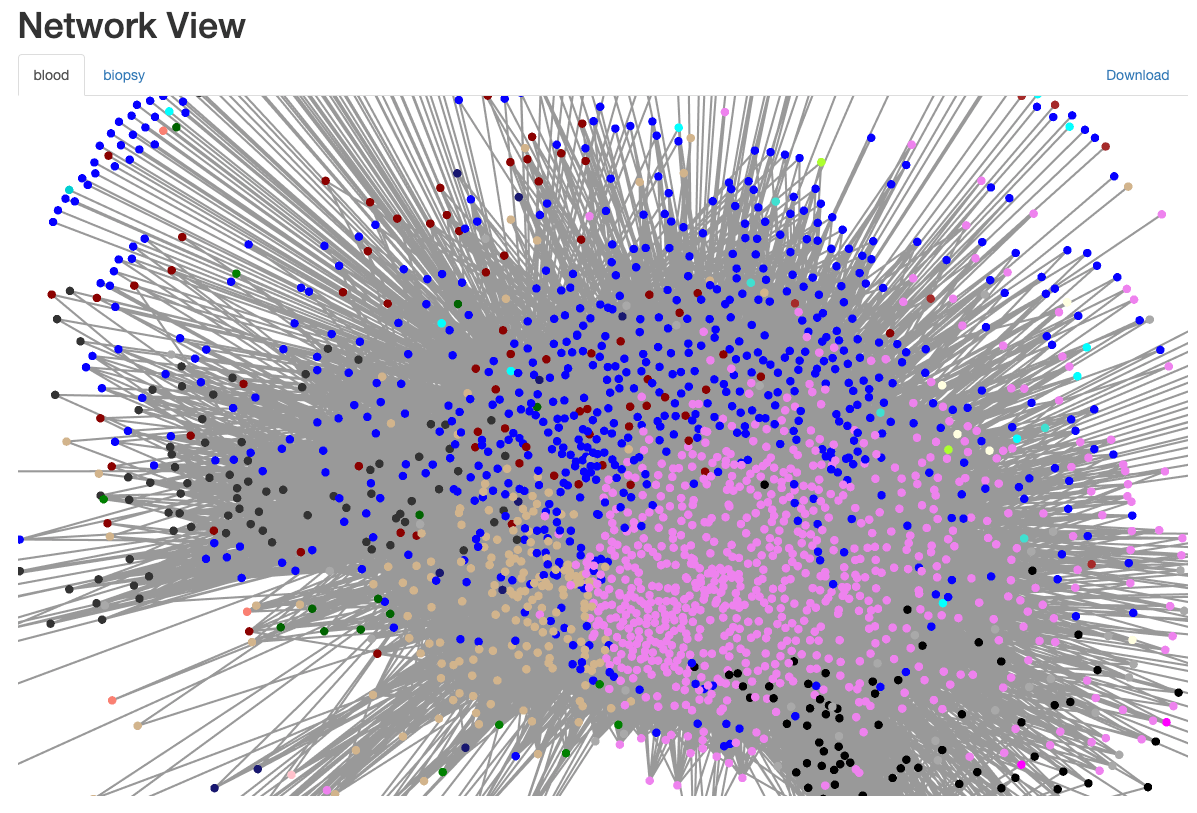}
\caption{Hairball problem in the network view in MIxT.}
\label{fig:hairball}
\end{figure}

The Matched Interaction Across Tissues (MIxT) is a tool for exploring and comparing transcriptional profiles from two or more matched tissues across individuals \cite{fjukstad_dumeaux_olsen_lund_hallett_bongo_2017}. The tool is implemented as a web application and it has a 2-dimensional visualization tool to explore biological networks\footnote{\url{https://mixt-tumor-stroma.bci.mcgill.ca/network}}. This tool has, however, some known scalability and visualization problems. When the user zooms in the network, we get the hairball problem as shown in figure \ref{fig:hairball}, since there are too many nodes and edges in the view. In addition, the edges need to be rendered every time the user interacts with the network, taking a few seconds to show all the visual information. We therefore use MIxT in GeneNet VR as a case study when designing and evaluating GeneNet VR to evaluate and demonstrate the benefits for VR in the interpretation of large biological networks .

MIxT visualizes two networks of gene-to-gene interactions, one for data from blood and the other from tissue. The networks are undirected graphs, so the nodes are connected with bidirectional edges. In a gene-to-gene interaction network the nodes represent genes and an edge between two nodes represents a significant co-expression relationship. The nodes in the networks are also organized in clusters and colors. Each cluster corresponds to a module, which is a subgraph where the genes are highly connected and where these genes are part of a common biological process that causes many interactions among themselves.

\section{GeneNet VR}

GeneNet VR is a virtual reality application for the interactive visualization of gene networks in a 3D space. To explore and visualize the data in GeneNet VR, the user can walk around the 3D environment, zoom in the network, translate (move from one place to another) the network, filter the nodes using a user interface, morph transition from one network to another and finally, also obtain detailed information about the nodes.

GeneNet VR loads node and relationship data from files resulting from bioinformatics analyses and then builds the network using a custom heuristic force-directed graph layout algorithm that distributes the nodes in the 3D space based on their co-expression. Finally, the user can explore it and interact with it using the VR headset and controllers.

We implemented GeneNet VR in C\# in Unity, a cross-platform game engine. Unity is used for a wide range of applications, especially for video games in 3D and 2D, VR applications, and civil engineering applications. We also used VRTK, a VR toolkit to build VR solutions in Unity. We used an Oculus Quest headset. It is an all-in-one head mounted display (HMD), which means that it doesn’t need to be connected to a PC. However, it can also be used with a PC with a powerful GPU if  more computing power than the standalone headset is needed.

\begin{figure}[h]
\includegraphics[width=0.7\textwidth]{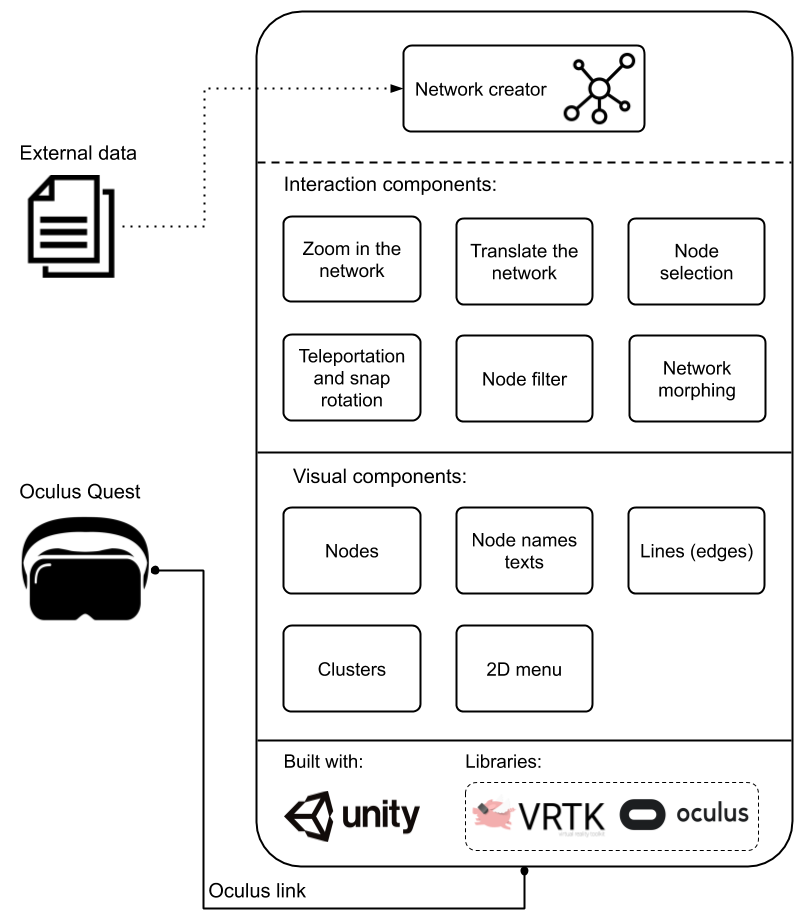}
\caption{Architecture and design of GeneNet VR}
\label{fig:architecture}
\end{figure}

\subsection{Architecture and Design}

GeneNet VR is a VR application built in Unity. For the implementation of the different visualization and interaction components we implmented C\# scripts, in addition to solutions that are native in Unity like the particle systems, and made use of the VRTK library and the Oculus library for Unity.

In Figure \ref{fig:architecture} we have an overview of the architecture of GeneNet VR. The big-box represents the Unity application and it contains all the components and functionalities that I have developed for the project. We can see that the big box is divided into 4 regions. The first one, starting from the top, is the network creator component that uses external data in order to build the network. In the second region, we have different interaction components that are available for the user to interact with the network and the environment. The third region contains the visual components that help the user visualize the data. Finally, the last region contains the technologies and libraries that we have used to build the application. We can also see the Oculus Quest headset represented down on the left. Here the user can visualize the network and use the controllers to interact. As we can see in the figure, the Oculus Quest can be connected to the PC using an Oculus Link, which is basically a high-quality USB 3 C to C or USB A to C cable with proven performance \cite{oculus_link}. This allows the user to run GeneNet VR on the PC. Another possibility is also to load GeneNet VR in the Oculus Quest and run it in the hardware of the headset without any cable or PC.

\subsection{Network visualization}

The network is created using data files that contain the network in CSV format. GeneNet VR stores the network in hash maps, using the gene name as key and a particle object as value, that can later be used by the interaction components to transform or read the data of the networks like the node positions or colors. The clustering algorithm is also applied when building the network, after which all particles are inserted into a Unity Particle System. 

The nodes are visualized as square boxes, with the color determined by the group assigned to each node in the data set. Edges are visualized as lines, signifying a relationship between two nodes that exceeds a threshold. The Names of nodes are displayed next to the selected node, as well as over the virtual controller when targeting a node. Clusters are not identified beyond the clustering algorithm used to group related nodes together, and therefore no visual identification of clusters is added beyond their spatial proximity. In addition, the menu appears in virtual space a fixed distance from the user when activated, after which the user can select items in the menu freely, as well as move around the menu.

\begin{figure}
    \begin{center}
    \includegraphics[width=0.7\textwidth]{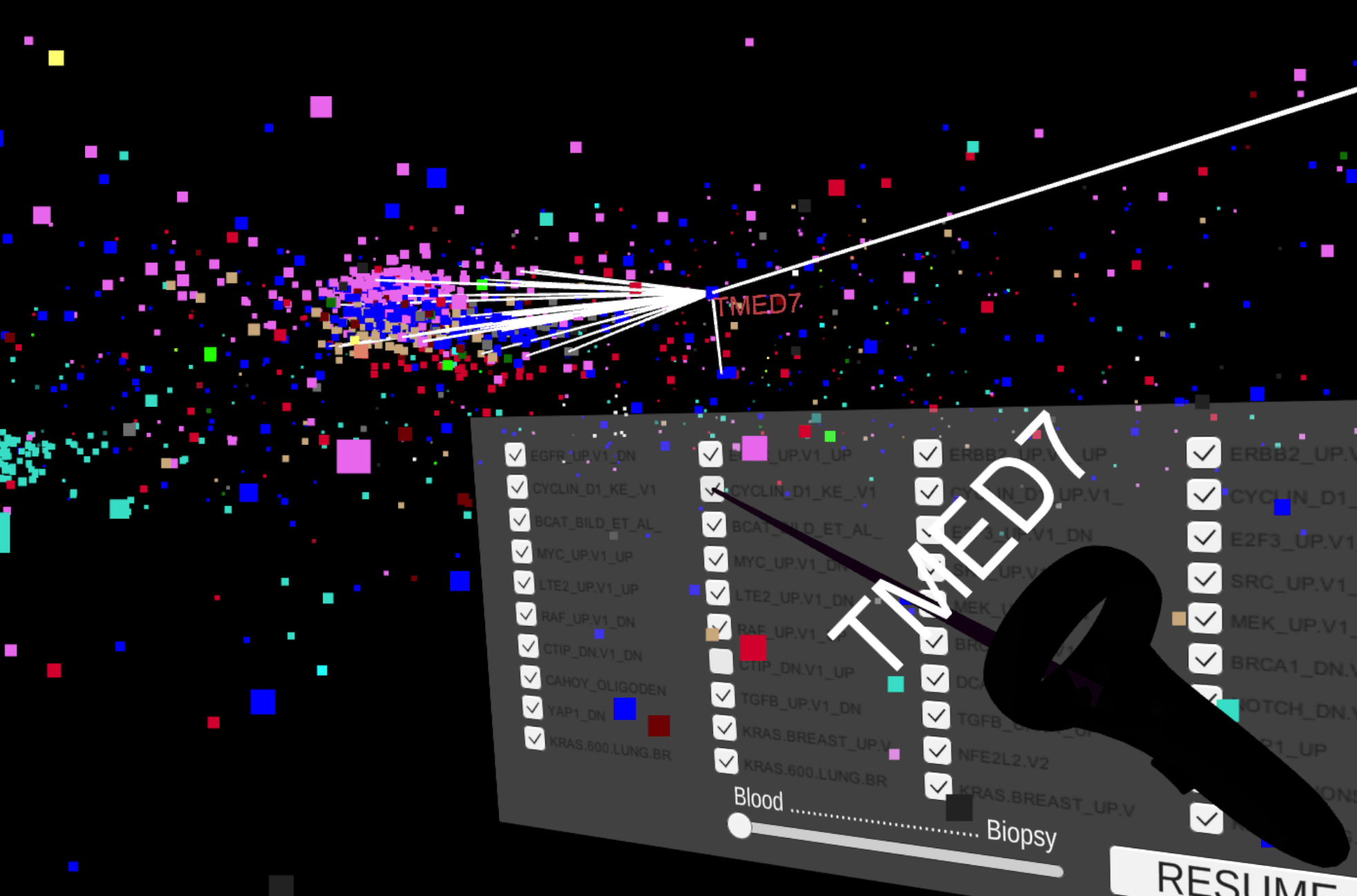}
    \caption{GeneNetVR user interface, showing the graph with a node selected, while using the controller to select an item in the menu.}
    \label{fig:interface}
    \end{center}
\end{figure}

\subsection{Interacting with a visualized network}

GeneNet VR provides interactions that allow the user to move both him/herself in small and large scales, as well as move and scale the network. In addition the user can select and filter the network nodes and morph the network to explore the data.

\subsubsection{View and locomotion}

The view in GeneNet VR is primarily controlled through the use of a 6 degrees-of-freedom headset, such that the user can look around, as well as move, with the movements in the real world mapped 1-to-1 to the visualization software. In addition to this, GeneNet VR supports teleportation through selecting a spot in the virtual world, as well as a rotation. The teleportation is performed with a “blink” effect to alleviate potential motion sickness. There is also the option to rotate the view in 45-degree increments to support tethered and sitting use. The user interaction used for teleportation is press-and-hold on a button, aim with the controller, and release to teleport. The interaction for rotation is moving the right thumbstick left or right.

\subsubsection{Translation and scaling}

In addition to moving the user’s viewpoint, it is also possible to translate and scale the network itself. These interactions support detailed inspection of e.g. dense clusters of nodes. Translation of the network is achieved by pressing and holding the “grip” button on the controller, and moving the controller. The network is moved in real-time, and movement continues until the grip button is released. Scaling and rotation are achieved by using both grip buttons, and moving the controllers in relation to each other.

\subsubsection{Selection and filtering}

Nodes are selected to provide more information about a single node, as well as visualizing the edges to other nodes. When a node is selected, edges from that node become visible. It is also possible to filter out classes of nodes, making them invisible and non-selectable. Selection is achieved through using the trigger button while aiming at a node. The controller has a laser pointer to assist with aiming at a node. Filtering is achieved through bringing up the menu by pressing the menu button, which brings up a menu where categories can be selected or deselected through the use of checkmark boxes.

\subsubsection{Morphing}

A key feature of the MiXT network is that there are two networks involved, where there may be some relation between some nodes across networks. To explore this, GeneNet VR supports morphing between the two networks. This is achieved through a slider in the menu, where the user brings up the menu with the menu key, and uses the slider to move from one network to the other. During the transition, the position and color of nodes are linearly interpolated between the networks. Nodes are only selectable when fully transitioned to one network or the other.

\section{Performance evaluation}

\subsection{Methodology}

To evaluate the performance of the MIxT prototype, we have implemented a benchmark that automatically performs a series of interactions: translating the network, scaling the network, and selecting nodes. These are the interactions that require the most computation and therefore represent the most challenging operations to achieve the required performance for. These interactions are performed over a span of 700 frames. The measurements start 500 frames into the benchmark to avoid cold-start issues. We record the time taken to render each frame. The frame times are evaluated in accordance with the Oculus Guidelines\cite{oculus_performance_baselines}, with a primary aim of maintaining a 72 FPS average, and avoiding stuttering due to long frame times for individual frames.

We use the “blood” data set from MiXT, which contains 2693 nodes and 89120 edges. Two subsets of this data set, with 2/3rd and 1/3rd of the original nodes and edges, are also used to evaluate scalability.

We report the average frame time of all, the 1\% slowest, and 0.25\% slowest frame rendering times. The 1\% and 0.25\% slowest measure potential stuttering during rendering, and therefore a bad experience for the user. Note that the longest possible frame time is 100 ms. If a frame is not updated within 100 ms that process is suspended and the next frame is  rendered

The benchmark has been performed on a desktop computer with an Intel Xeon E3-1275 V6 CPU, 64 GB RAM, and an NVIDIA GeForce GTX 2080 Ti GPU. The experiments were repeated on an Oculus Quest, which uses a Qualcomm Snapdragon 835 SoC with a Qualcomm Adreno 540 GPU solution.

\subsection{Results}

\begin{figure}
    \begin{minipage}[c]{0.45\linewidth}
        \includegraphics[width=\textwidth]{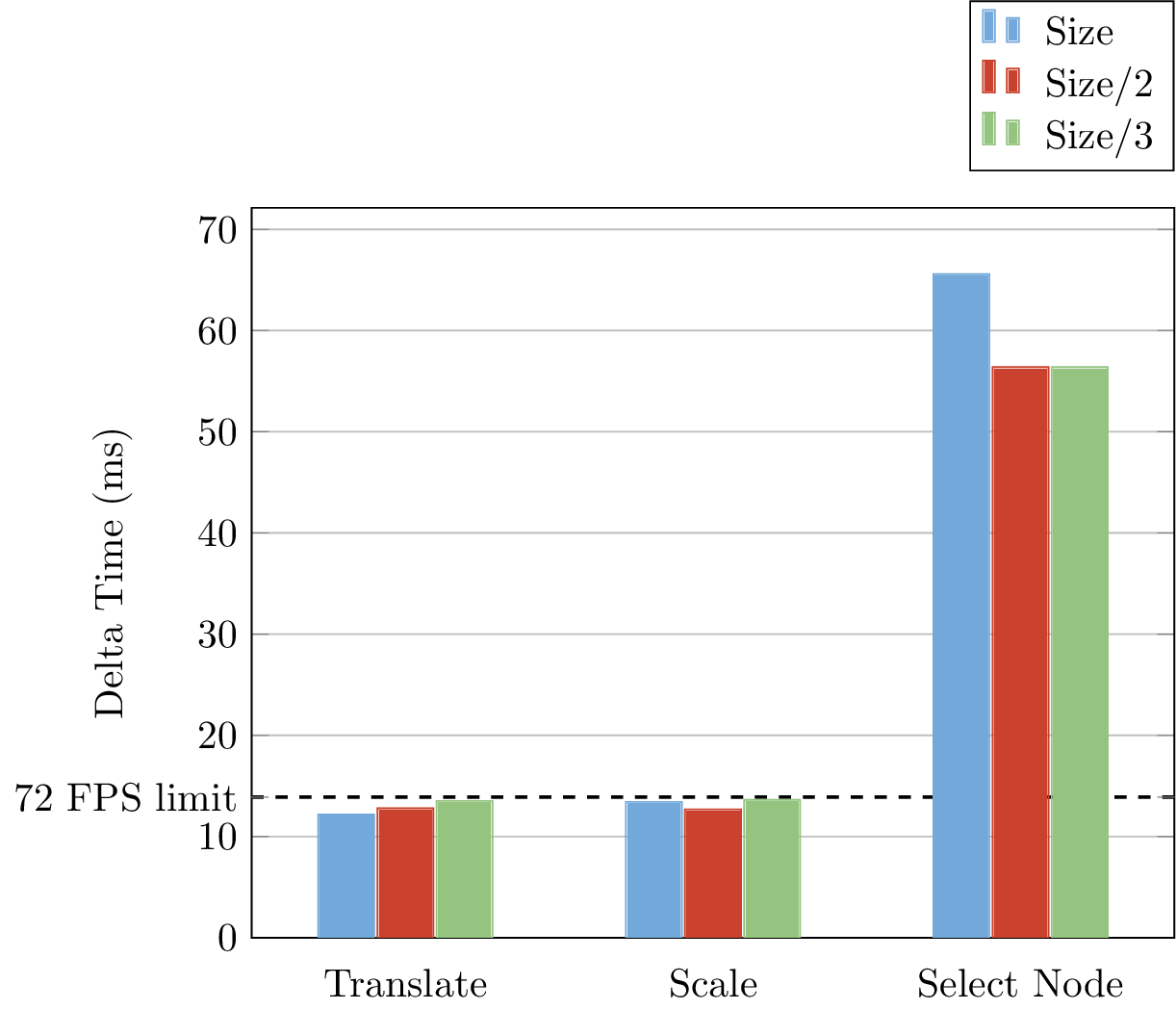}
        \caption{Comparison of the 1\% slowest frames to the recommended frames per second.}
        \label{fig:performance}
    \end{minipage}
    \begin{minipage}[c]{0.45\linewidth}
        \includegraphics[width=\textwidth]{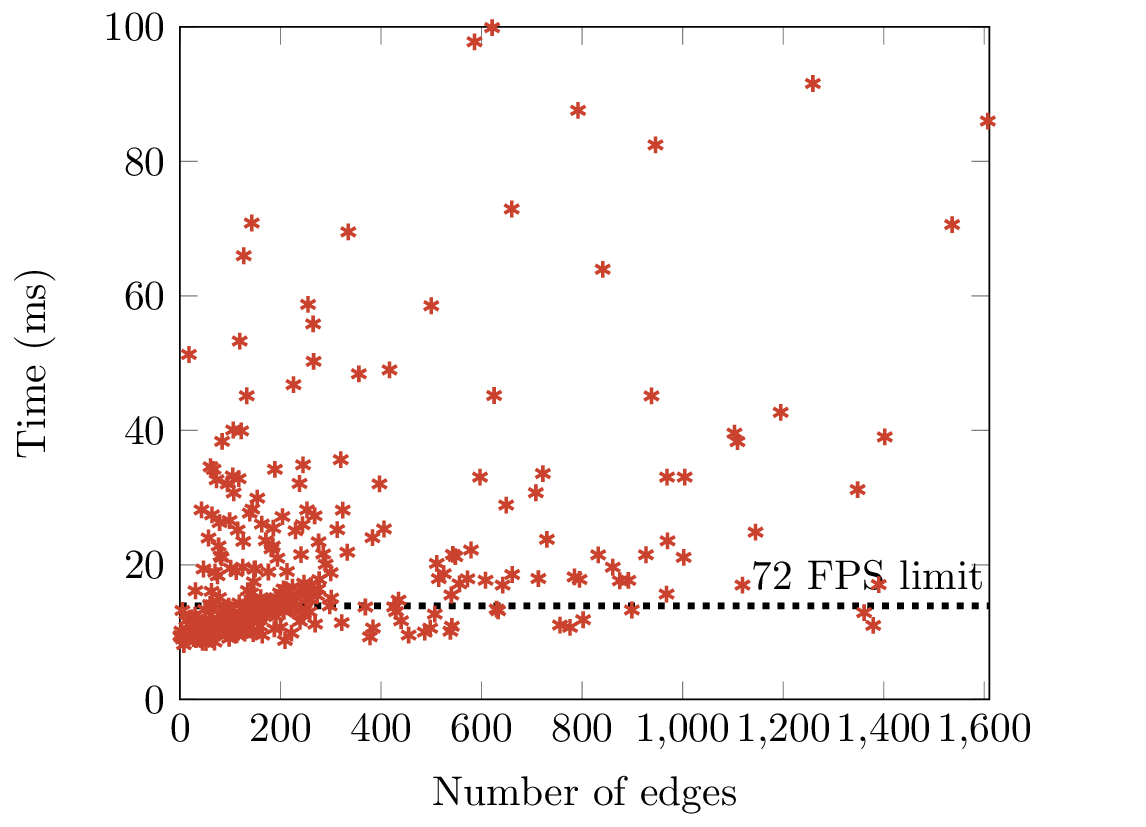}
        \caption{Relationship between the number of edges for a node, and how many milliseconds of latency are introduced when selecting the node.}
        \label{fig:scatter}
    \end{minipage}
\end{figure}

The results, shown in table \ref{tab:performance}, show that translation and scaling operations are within the Oculus Guidelines, with very few frames  exceeding the recommended 14 milliseconds per frame (72 FPS). However, the selection operation takes longer for at least 1\% of the frames (figure \ref{fig:performance}). These results indicate that stuttering may be experienced when selecting nodes, while other operations operate within the guidelines.

Profiling the selection interaction shows a weak relationship between frame time and number of edges (figure \ref{fig:scatter}). In the cases where there are long frame times, the time is spent in object management, indicating that there is an issue in our implementation of generating edges for visualization. We believe that this issue can be resolved without impacting the overall design.

\begin{table}[]
    \centering
    \begin{tabular}{l|l|l|l}
        \hline
         Interaction & Avg. of 0.25\% slowest & Avg. of 1\% slowest & Avg. of all \\
         \hline
         Translation full & 20.11 & 12.17 & 6.55 \\
         Translation 1/3rd & 23.28 & 13.52 & 6.5 \\
         Scaling full & 22.82 & 13.41 & 6.51 \\
         Scaling 1/3rd & 23.02 & 13.63 & 6.49 \\
         Select full & 100 & 65.57 & 8.71 \\
         Select 1/3rd & 100 & 56.39 & 8.58 \\
    \end{tabular}
    \caption{Summary of performance results, all times reported in milliseconds}
    \label{tab:performance}
\end{table}

\begin{figure}
\includegraphics[width=0.5\textwidth]{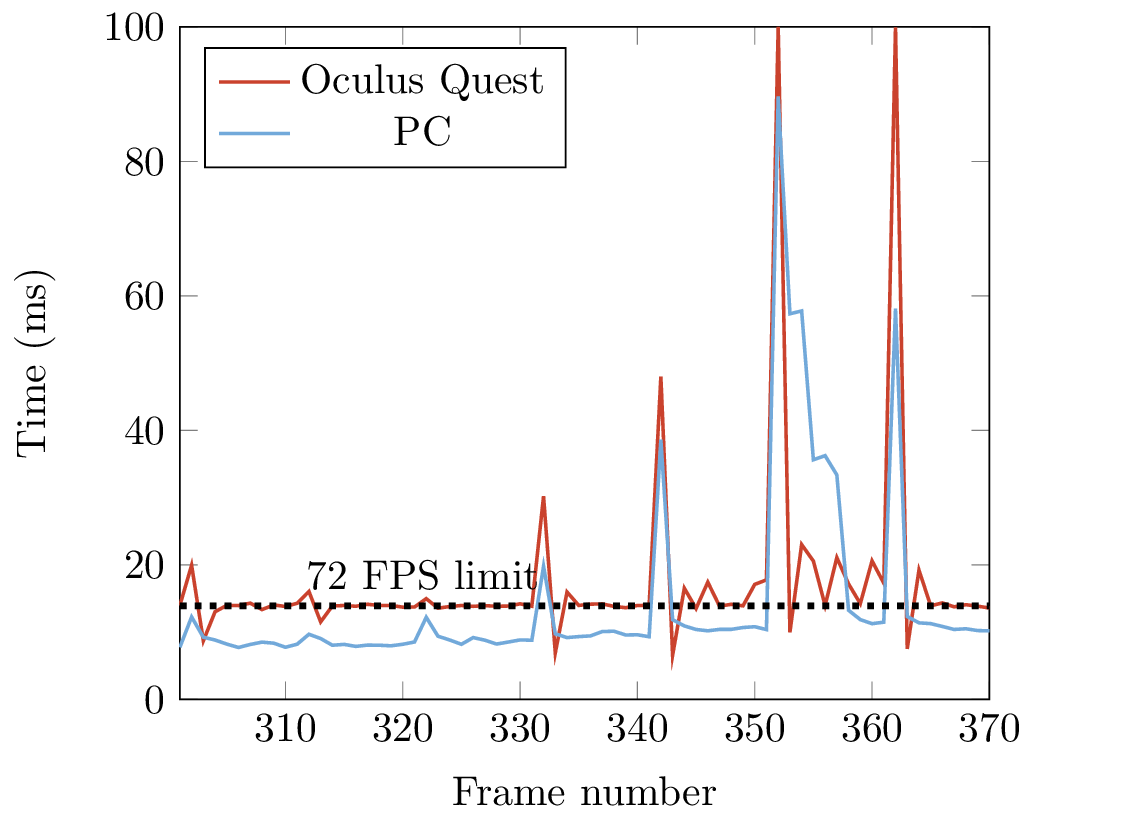}
\caption{Plot of frame times on the Oculus Quest standalone and a PC while performing the benchmark.}
\label{fig:standalone}
\end{figure}

For the MIxT dataset we found that the performance on Quest standalone is similar to performance on an  external GPU, see figure \ref{fig:standalone}. These results show that using the Oculus Quest in standalone mode is a viable choice for this use case.

\section{Usability evaluation}
\subsection{Methodology}

To evaluate the usability and user experience of GeneNet VR for visualizing large-scale biological networks, we interviewed six active researchers within related fields at UiT The Arctic University of Norway (Table \ref{tab:experience}). The interviews were semi-structured with three open-ended questions:
\begin{enumerate}
    \item How do you perceive the application?
    \item How do you perceive the application for pattern finding?
    \item What is missing in the application?
\end{enumerate}

Before starting the interview, a preliminary introduction to each respondent was given. We explained the research problem, as well as what MIxT is, the datasets that we are visualizing, and how they can use the different interactions with the VR controllers in GeneNet VR. Then the respondents tested the application using the Oculus Quest connected to the machine. We discussed the three questions, and recorded the discussion by taking notes. The notes were later analyzed.

\begin{table}[]
    \centering
    \begin{tabular}{l|l|l}
        \hline
         Respondent & Research field & VR experience \\
         \hline
         1 & Computer science. Has  & Owns a HTC Vive and  \\
           & worked with social networks. & uses it on a regular basis \\ 
           \hline
         2 & Clinical pharmacy and & Has used VR several  \\
           & pharmacoepidemiology & times to play video games \\
           \hline
         3 & Pharmacoepidemiology. Has worked  & Never used \\
           & with drug interaction networks & \\
           \hline
         4 & Biology. Has worked with & Very little \\
           & gene co-expression networks & \\
           \hline
         5 & Biology. Has worked with & Never used \\
           & gene expression analysis & \\
           \hline
         6 & Biology. Has worked with & Never used \\
           & GCN and biological networks & \\
           \hline
    \end{tabular}
    \caption{Respondent background and VR experience}
    \label{tab:experience}
\end{table}

\subsection{Results}

\subsubsection{Perception of the application}
All participants agreed that the use of GeneNet VR had some advantages when compared to traditional visualization techniques. None of the respondents felt that the performance negatively impacted their experience. A bug with the node selection required a restart of the application two times.

Some guidance was needed for the interaction for the users who had not used VR before, while the users that had used VR before required very little guidance. Respondents 4, 5, and 6 highlighted that the interactions are intuitive and easy to learn. Respondents 1 and 5 felt that using a standalone headset like the Oculus Quest was advantageous. Respondents 2 and 3 preferred to use this kind of visualization tool while sitting. This shows that the choice of interaction methods were appropriate, some guidance is needed for novice users, and the potential stuttering when selecting nodes did not impact the experience negatively.

\subsubsection{Pattern finding}
All respondents expressed that finding patterns in GeneNet VR seemed to be easier than in other visualization tools they have used. Respondents 1, 2, 3 and 4 said that GeneNet VR would be helpful to find patterns in their own data sets. Participant 4 was particularly interested in using GeneNet VR on a data set they were working on currently. The participants with a background in pharmacoepidemiology remarked on the similarities between their drug data sets and the MiXT data set. This shows that visualizing this category of data in VR is attractive, and that users are eager to explore their data in VR.

\subsubsection{Improvements}
During the discussions, the following key areas of improvement were highlighted:
\begin{enumerate}
    \item We should show the strength of co-expression using for example line thickness or line color. The distance between one and another node can also give important information about this. 
    \item We should show the names of connected nodes when a node is selected.
    \item It should be possible to rotate the network.
    \item It should be possible to change the layout of the network
    \item It should be possible to filter the nodes by clusters.
    \item There should be a way to search for node functionality.
\end{enumerate}

Many other improvements were also discussed, and a complete listing is available in \cite{alvaro}.

\section{Related work}

Virtual Reality Chemical Space is a VR application for the interactive exploration of chemical space populated by Drugbank compounds\cite{drugbank}. It is also developed in Unity using C\# and the VRTK library. They use a particle system to render the particles of the chemical space. To render the particles, they use shaders instead of geometrical spheres as well, optimizing the number of vertices per datapoint in the scene. In order to reduce motion sickness, they have introduced a floor in the form of a grid acting as a static frame of reference. Also instead of letting the user move through the VR environment, they only use a controller to move the point cloud.

BioVR is an interactive VR platform for integrated visual analysis of DNA/RNA protein structures\cite{biovr}. It is built in Unity and using C\#. The headset that they targeted the application to is Oculus Rift. One big difference between BioVR and our application is that in BioVR they use the hands for the interactions rather than the controllers.

CellexalVR is a virtual reality environment for the visualization and analysis of single-cell RNAseq experiments that help researchers understand their data\cite{cellexalvr}. The system is divided into two parts: the first one consists of the VR interface and the second is an R package called cellexalvrR that does back- end calculations and also provides functions that allows the user to export the scRNAseq data from an R session for CellexalVR to read. CellexalVR was developed in Unity for HTC Vive (Pro). They used Unity and C\# and R for the implementation. They used libraries like VRTK, OpenVR, and SteamVR as well in the implementation.

BigTop is a visualization framework in VR for the rendering of Manhattan plots in three dimensions\cite{bigtop}. For the interaction, BigTop allows the user to select a node in order to obtain more information like the SNP name. BigTop is built in JavaScript with the React and A-Frame frameworks. It can also be rendered in any commercially available VR headsets and also in 2D web browsers.

\section{Discussion}

We ran several performance experiments for our case study and demonstrated that GeneNet VR performs well when exploring the networks from MIxT. We also evaluated the performance on the machine for several interactions that are commonly used during the visualization process and obtained an average of 7-8 milliseconds, which is under the 13.9 milliseconds limit that corresponds to 72 FPS required by Oculus. Likewise, we evaluated the performance on the Oculus Quest hardware and the results indicated that the Oculus Quest is around 30\% slower than on the PC, but it still reaches the 72 FPS goal.

The performance evaluation shows that selection can incur a stutter when the edges for a node are generated. However, the users interviewed did not find this to be a significant detractor from the usability of the application.

The interviews revealed that the use of VR for exploring this type of data may be beneficial for finding patterns, and that little training is required for users. In addition, this type of visualization appears to be relevant for several different fields of study.

Our work has some two main limitations. First, although the two MIxT networks used in our evaluation are large, there are other biological visualizations with much larger networks. We believe the MIxT networks were not large enough to hit the performance ceiling on the Oculus Quest. Further experiments are therefore required with even larger networks to better understand the performance limitations of standalone and GPU accelerated VR network visualizations and interactions. Second, we have not implemented and used the visualized networks for real analyses. Additional interaction patterns, network manipulation methods may therefore be required. The user interviews identified some of the needed functionality. However, we believe the implemented methods show that these can be implemented and achieve the required performance for a smooth VR experience. 

\section{Conclusion}

In this paper we have implemented GeneNet VR, a system for visualizing biological graph data sets. This system has been implemented on a standalone VR headset, and has been evaluated through a performance evaluation, as well as interviews with potential users. Our evaluation shows that this approach is attractive for users, as well as technologically feasible for real-scale data sets. However, additional work is required to evaluate its benefits to improve knowledge discovery for real data analysis use cases.  

GeneNet VR is open-sourced: \url{https://github.com/kolibrid/GeneNet-VR}. 
A video demonstrating GeneNet VR used to explore large biological networks: \url{https://youtu.be/N4QDZiZqVNY}.

\section{Acknowledgements}
Thanks to Rafael Nozal Cañadas and Nikita Shvetsov for their comments to the paper, and the researchers participating in the usability evaluation.

\bibliographystyle{splncs04}
\bibliography{paper}

\end{document}